\documentclass{webofc}
\usepackage[varg]{txfonts}   
\usepackage{lipsum}
\begin{document}
\title{The NIKA2 Sunyaev-Zeldovich Large Program}
\subtitle{Precise galaxy cluster physics for an accurate cluster-based cosmology}
\author{\firstname{L.}~\lastname{Perotto}\inst{\ref{LPSC}}\fnsep\thanks{\email{laurence.perotto@lpsc.in2p3.fr}}
  \and \firstname{R.}~\lastname{Adam} \inst{\ref{LLR}}
  \and  \firstname{P.}~\lastname{Ade} \inst{\ref{Cardiff}}
  \and  \firstname{H.}~\lastname{Ajeddig} \inst{\ref{CEA}}
  \and  \firstname{P.}~\lastname{Andr\'e} \inst{\ref{CEA}}
  \and \firstname{M.}~\lastname{Arnaud} \inst{\ref{CEA}}
  \and \firstname{E.}~\lastname{Artis} \inst{\ref{LPSC}}
  \and  \firstname{H.}~\lastname{Aussel} \inst{\ref{CEA}}
  \and  \firstname{I.}~\lastname{Bartalucci} \inst{\ref{Milano}} 
  \and  \firstname{A.}~\lastname{Beelen} \inst{\ref{IAS}}
  \and  \firstname{A.}~\lastname{Beno\^it} \inst{\ref{Neel}}
  \and  \firstname{S.}~\lastname{Berta} \inst{\ref{IRAMF}}
  \and  \firstname{L.}~\lastname{Bing} \inst{\ref{LAM}}
  \and  \firstname{O.}~\lastname{Bourrion} \inst{\ref{LPSC}}
  \and  \firstname{M.}~\lastname{Calvo} \inst{\ref{Neel}}
  \and  \firstname{A.}~\lastname{Catalano} \inst{\ref{LPSC}}
  \and  \firstname{M.}~\lastname{De~Petris} \inst{\ref{Roma}}
  \and  \firstname{F.-X.}~\lastname{D\'esert} \inst{\ref{IPAG}}
  \and  \firstname{S.}~\lastname{Doyle} \inst{\ref{Cardiff}}
  \and  \firstname{E.~F.~C.}~\lastname{Driessen} \inst{\ref{IRAMF}}
  \and  \firstname{A.}~\lastname{Ferragamo} \inst{\ref{Roma}}
  \and  \firstname{A.}~\lastname{Gomez} \inst{\ref{CAB}}
  \and  \firstname{J.}~\lastname{Goupy} \inst{\ref{Neel}}
  \and  \firstname{F.}~\lastname{K\'eruzor\'e} \inst{\ref{LPSC}}
  \and  \firstname{C.}~\lastname{Kramer} \inst{\ref{IRAME}}
  \and  \firstname{B.}~\lastname{Ladjelate} \inst{\ref{IRAME}}
  \and  \firstname{G.}~\lastname{Lagache} \inst{\ref{LAM}}
  \and  \firstname{S.}~\lastname{Leclercq} \inst{\ref{IRAMF}}
  \and  \firstname{J.-F.}~\lastname{Lestrade} \inst{\ref{LERMA}}
  \and  \firstname{J.-F.}~\lastname{Mac\'ias-P\'erez} \inst{\ref{LPSC}}
  \and  \firstname{A.}~\lastname{Maury} \inst{\ref{CEA}}
  \and  \firstname{P.}~\lastname{Mauskopf} \inst{\ref{Cardiff},\ref{Arizona}}
  \and \firstname{F.}~\lastname{Mayet} \inst{\ref{LPSC}}
  \and  \firstname{A.}~\lastname{Monfardini} \inst{\ref{Neel}}
  \and  \firstname{M.}~\lastname{Mu\~noz-Echeverr\'ia} \inst{\ref{LPSC}}
  \and  \firstname{A.}~\lastname{Paliwal} \inst{\ref{Roma}}
  \and  \firstname{G.}~\lastname{Pisano} \inst{\ref{Cardiff}}
  \and  \firstname{E.}~\lastname{Pointecouteau} \inst{\ref{Toulouse}}
  \and  \firstname{N.}~\lastname{Ponthieu} \inst{\ref{IPAG}}
  \and  \firstname{G.~W.}~\lastname{Pratt} \inst{\ref{CEA}}
  \and  \firstname{V.}~\lastname{Rev\'eret} \inst{\ref{CEA}}
  \and  \firstname{A.~J.}~\lastname{Rigby} \inst{\ref{Cardiff}}
  \and  \firstname{A.}~\lastname{Ritacco} \inst{\ref{IAS}, \ref{ENS}}
  \and  \firstname{C.}~\lastname{Romero} \inst{\ref{Pennsylvanie}}
  \and  \firstname{H.}~\lastname{Roussel} \inst{\ref{IAP}}
  \and  \firstname{F.}~\lastname{Ruppin} \inst{\ref{MIT}}
  \and  \firstname{K.}~\lastname{Schuster} \inst{\ref{IRAMF}}
  \and  \firstname{S.}~\lastname{Shu} \inst{\ref{Caltech}}
  \and  \firstname{A.}~\lastname{Sievers} \inst{\ref{IRAME}}
  \and  \firstname{C.}~\lastname{Tucker} \inst{\ref{Cardiff}}
  \and  \firstname{G.}~\lastname{Yepes} \inst{\ref{Madrid}}
}
  
  \institute{
    Univ. Grenoble Alpes, CNRS, Grenoble INP, LPSC-IN2P3, 38000 Grenoble, France
    \label{LPSC}
    \and
    LLR, CNRS, École Polytechnique, Institut Polytechnique de Paris, Palaiseau, France
    \label{LLR}
    \and
    School of Physics and Astronomy, Cardiff University, Queen’s Buildings, The Parade, Cardiff, CF24 3AA, UK 
    \label{Cardiff}
    \and
    AIM, CEA, CNRS, Universit\'e Paris-Saclay, Universit\'e Paris Diderot, Sorbonne Paris Cit\'e, 91191 Gif-sur-Yvette, France
    \label{CEA}
    \and
    INAF, IASF-Milano, Via A. Corti 12, 20133 Milano, Italy
    \label{Milano}            
    \and
    Institut d'Astrophysique Spatiale (IAS), CNRS, Universit\'e Paris Sud, Orsay, France
    \label{IAS}
    \and
    Institut N\'eel, CNRS, Universit\'e Grenoble Alpes, France
    \label{Neel}
    \and
    Institut de RadioAstronomie Millim\'etrique (IRAM), Grenoble, France
    \label{IRAMF}
    \and
    Aix Marseille Univ, CNRS, CNES, LAM, Marseille, France
    \label{LAM}
    \and 
    Dipartimento di Fisica, Sapienza Universit\`a di Roma, Piazzale Aldo Moro 5, I-00185 Roma, Italy
    \label{Roma}
    \and
    Univ. Grenoble Alpes, CNRS, IPAG, 38000 Grenoble, France 
    \label{IPAG}   
    \and
    Centro de Astrobiolog\'ia (CSIC-INTA), Torrej\'on de Ardoz, 28850 Madrid, Spain
    \label{CAB}
    \and  
    Instituto de Radioastronom\'ia Milim\'etrica (IRAM), Granada, Spain
    \label{IRAME}
    \and 
    LERMA, Observatoire de Paris, PSL Research University, CNRS, Sorbonne Universit\'e, UPMC, 75014 Paris, France  
    \label{LERMA}
    \and
    Univ. de Toulouse, UPS-OMP, CNRS, IRAP, 31028 Toulouse, France
    \label{Toulouse}
    \and 
    Laboratoire de Physique de l’\'Ecole Normale Sup\'erieure, ENS, PSL Research University, CNRS, Sorbonne Universit\'e, Universit\'e de Paris, 75005 Paris, France
    \label{ENS}
    \and
    Department of Physics and Astronomy, University of Pennsylvania, 209 South 33rd Street, Philadelphia, PA, 19104, USA
    \label{Pennsylvanie}   
    \and 
    Institut d'Astrophysique de Paris, CNRS (UMR7095), 98 bis boulevard Arago, 75014 Paris, France
    \label{IAP}
    \and 
    Kavli Institute for Astrophysics and Space Research, Massachusetts Institute of Technology, Cambridge, MA 02139, USA
    \label{MIT}
    \and
    School of Earth and Space Exploration and Department of Physics, Arizona State University, Tempe, AZ 85287, USA
    \label{Arizona}
    \and
    Caltech, Pasadena, CA 91125, USA
    \label{Caltech}
    \and
    Departamento de F\'isica Te\'orica and CIAFF, Facultad de Ciencias, Modulo 8, Universidad Aut\'anoma de Madrid, 28049 Madrid, Spain
    \label{Madrid}
  }

  \abstract{%
    The NIKA2 Guaranteed-Time SZ Large Program (LPSZ) is dedicated to the high-angular resolution SZ mapping of a representative sample of 45 SZ-selected galaxy clusters drawn from the catalogues of the Planck satellite, or of the Atacama Cosmology Telescope. The LPSZ sample spans a mass range from $3$ to $11 \times 10^{14} M_{\odot}$ and a redshift range from $0.5$ to $0.9$, extending to higher redshift and lower mass the previous samples dedicated to the cluster mass calibration and universal properties estimation. The main goals of the LPSZ are the measurement of the average radial profile of the ICM pressure up to $R_{500}$ by combining NIKA2 with Planck or ACT data, and the estimation of the scaling law between the SZ observable and the mass using NIKA2, XMM-Newton and Planck/ACT data. Furthermore, combining LPSZ data with existing or forthcoming public data in lensing, optical/NIR or radio domains, we will build a consistent picture of the cluster physics and further gain knowledge on the mass estimate as a function of the cluster morphology and dynamical state.

    We give an overview of the LPSZ, present recent results and discuss the future implication for cosmology with galaxy clusters.
}
\maketitle
\section{Introduction}
\label{intro}
Clusters of galaxies are powerful cosmological probes of the large-scale structure growth and of the Universe expansion over cosmic time~(see e.~g.~\cite{revue}). 
However, galaxy cluster-based cosmology is currently limited by the accuracy with which the mass of these objects can be inferred~\cite{Pratt2019}.
Galaxy clusters being scale-invariant, scaling laws that relate their mass to an observable can be built from representative sub-samples of clusters. In particular, from X-ray photometric and spectroscopic observation towards clusters, the spherical profiles of both the density and the temperature of the Intra-Cluster Medium (ICM) can be reconstructed, enabling the estimation of the mass under the hydrostatic equilibrium assumption. The main drawback of this method lies in the high-exposure time needed to achieve the data quality that permits an accurate estimate of the temperature profile, which become prohibitive for faint or distant clusters. As a consequence, widely-used reference samples, from which the mass-observable relation and universal properties of clusters are measured, are limited to massive and low-redshift clusters~\cite{Pratt2009, Planck2013}.

With the advent of high-angular-resolution millimeter-domain experiments, such as NIKA2, resolved observations of the Sunyaev-Zel'dovich effect~(SZ~\cite{SZ}) towards clusters offer a way of improving the accuracy and extending the coverage in mass and redshift of the relation between the hydrostatic mass and the integrated Compton parameter. Such observations also enable measuring the mean pressure profile of clusters. The SZ effect being independent on the redshift and proportional to the line-of-sight integral of the pressure of the ICM, accurate pressure profiles can be inferred in clusters up to high-redshift. Pressure profiles drawn from resolved SZ observations can then be combined with density profiles extracted from photometric X-ray observations to derive a hydrostatic mass estimate. Moreover, the mean pressure profile of clusters can be estimated in cluster samples from a direct observable of the pressure.

NIKA2 is a millimetre-domain camera installed at the IRAM 30-m telescope and opened to the community since 2017~\cite{NIKA2-general, NIKA2-instrument, NIKA2-electronics}. It operates in two frequency bands centered at $150$ and $260$~GHz, with an angular resolution of $17.6$ and $11.1''$, respectively, and offers high mapping speeds~\cite{NIKA2-performance}. With an angular resolution comparable to that of the XMM-Newton space observatory, NIKA2 will therefore enable the full deployment of the combined SZ and X-ray methods to infer the hydrostatic mass, and unveil the pressure distribution of the ICM. 
The NIKA2 SZ Large Program (LPSZ) consists of 300 hours of Guaranteed-Time dedicated to this goal.

%
\section{Overview of the LPSZ}
\label{Sect_LPSZ}

The LPSZ program relies on the state-of-the-art performance of the NIKA2 camera for the high-angular resolution mapping of the thermal SZ effect in a representative sample of galaxy clusters in view of improving the accuracy of cluster-based cosmology (P.I.: F.~Mayet; co-P.I.: L.~Perotto). 

\subsection{Sample}
\label{Sect_LPSZ_sample}

\begin{figure*}
\centering
\includegraphics[width=0.45\textwidth]{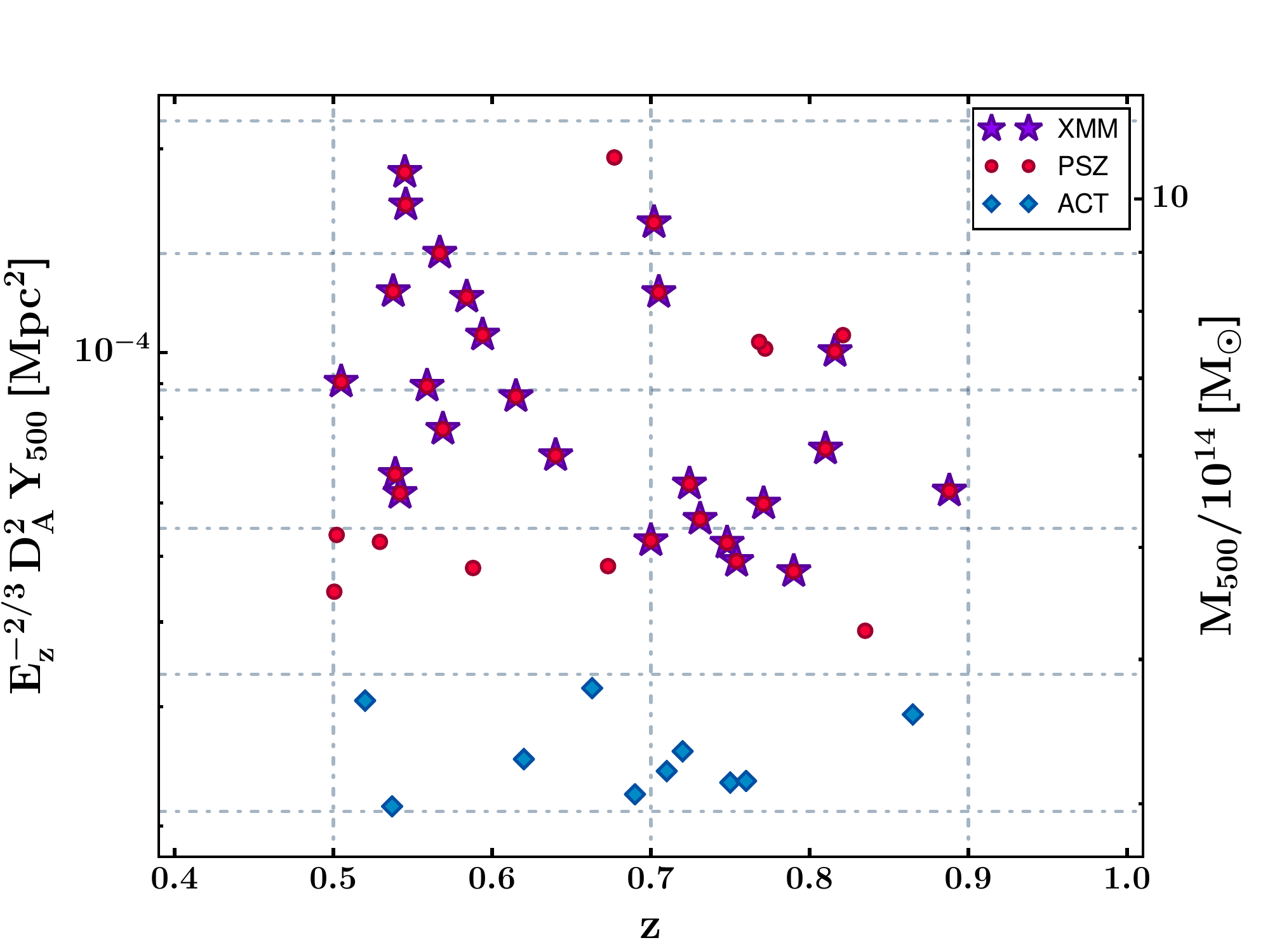}
\includegraphics[width=0.48\textwidth]{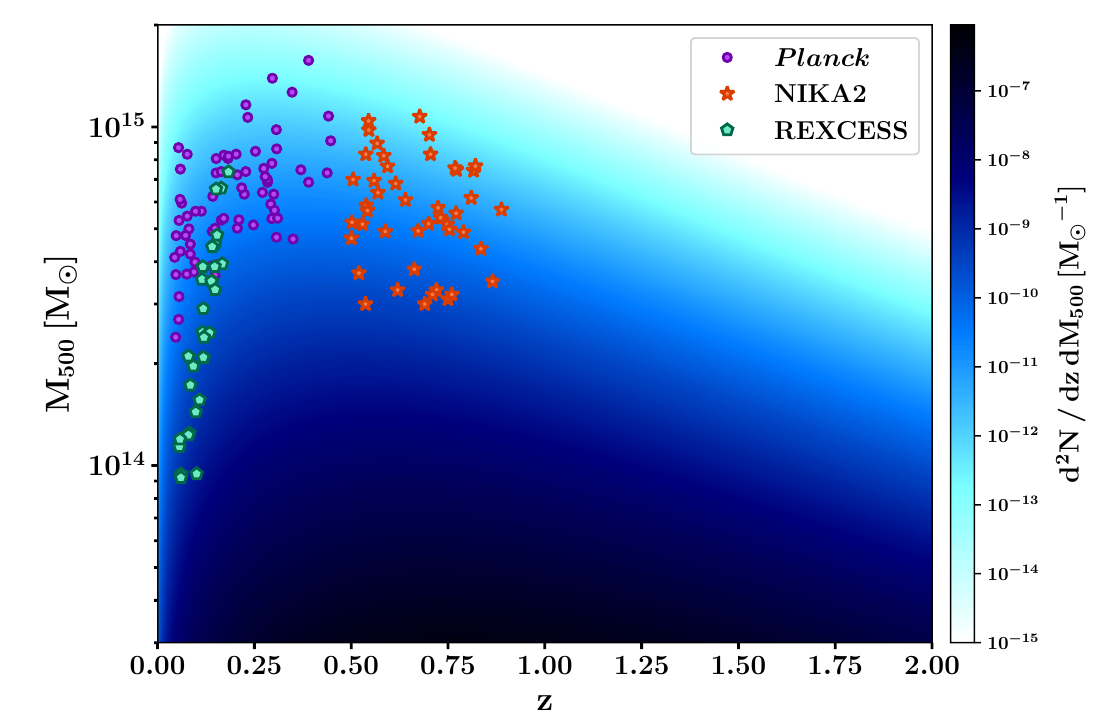}
\caption{\emph{Left:} Distribution of the LPSZ cluster sample in the plane of the redshift (abscissa) and the integrated Compton parameter (ordinate). The corresponding mass, as derived from the Planck Y-M scaling relation is given in the right ordinate-scale. The sample comprizes clusters drawn from the Planck (red) and ACT (blue) catalogs~\cite{PSZ2, ACT2013}, and already observed by XMM-Newton (stars).  \emph{Right:} Comparison of the LPSZ sample (red) with previous samples dedicated to the cluster mass calibration~\cite{Planck2013, Pratt2009} in the mass-redshift plane color-coded as a function of the predicted cluster abundance. Figure extracted from~\cite{Ruppin2019}.}
\label{Fig_LPSZ}       
\end{figure*}

The LPSZ sample consists of 45 galaxy clusters selected in intervals of mass and redshift in the ACT~\cite{ACT2013} and Planck~\cite{PSZ2} cluster catalogs. Being SZ-selected, the sample is representative of the whole cluster population. It covers intermediate to high redshifts (0.5 < z < 0.9) and spans a large range of masses from $3$ to $11 \times 10^{14} M_{\odot}$. The distribution of the sample as a function of the integrated Compton parameter and the redshift, as well as the mass and redshift bins defined for the sample selection, are shown in the left panel of Fig.~\ref{Fig_LPSZ}. Most of the targeted Planck clusters were observed in the X-ray domain with XMM-Newton or Chandra. X-ray observations of the Planck and ACT LPSZ clusters will be completed by the end of the program.

\subsection{Main Goals}
\label{Sect_LPSZ_goals}

The main goal of the LPSZ is providing the community with tools to improve the accuracy of the Cosmology with galaxy clusters, with a focus on the estimation of the mass. Combining SZ and X-ray observations at a similar angular resolution, we will perform in-depth characterizations of the thermodynamical properties of the ICM. For each cluster, we will provide spherical profiles of the ICM pressure, density, hydrostatic mass, entropy and temperature, independently of any X-ray spectroscopic data. From the complete sample, we will estimate the mean pressure profile of the clusters using SZ data only. In this endavour, we will benefit from NIKA2 unique combination of an angular resolution better than 18'' at 150~GHZ and a 6.5' field-of-view to extend previous studies in spanning a radial range from the core to the typical radius of the clusters. Moreover we will release a scaling relation between the integrated Compton parameter and the hydrostatic mass extending to higher redshift and lower mass previous estimates~\cite{Planck2013, Pratt2009}, as illustrated in the right panel of Fig.~\ref{Fig_LPSZ}.

\subsection{Complementary projects}
\label{Sect_LPSZ_complemet}


Several multi-wavelength analyses will be conducted. First of all, combining the integrated line-of-sight mass density, which can be reconstructed from lensing data, with the LPSZ estimates of the hydrostatic mass, we will study the lensing-to-hydrostatic-mass bias as a function of the cluster morphology and dynamical state. The intersection of the LPSZ and CLASH~\cite{CLASH} samples offers us a well-suited test bench of these studies~\cite{Miren}. Moreover about 50\% of the LPSZ sample have deep-integration optical data (SDSS, PanStarrs)~\cite{vanderBurg2018}, allowing us to explore the connexions between the hydrostatic mass and the richness or the dynamical mass estimates. Finally, we will conduct dedicated studies of the non-thermal properties of the ICM to gain knowledge on turbulences, AGN feedback or shocks, using e.~g.~cross-correlation with radio-domain data.

We will use state-of-the-art hydrodynamical simulations, such as the Three-Hundred project~\cite{Cui2018}, to select a series of synthetic samples of the same statistical or morphological properties as the LPSZ sample, and coming along with mock observations in SZ, X-ray and optical. These twin samples will be key to test novel methods, to assess the impact of dynamical state on cluster properties, to forecast the LPSZ sample sensitivity to redshift or mass evolution, to obtain accurate uncertainties on scaling laws and biases and to explore the LPSZ sample capability to measure the gas fraction or the Hubble parameter.

\section{Status of the observation and the analysis}
\label{Sect_status}
%

In order to achieve our main goals, the representativity of the sample must be preserved at the level of both the observation and the analysis. 
The observation time allocated to a cluster is defined so as to ensure a $3\sigma$ measurement of the Compton parameter profile at the angle $R_{500}/D_A$. Quantitatively, about half of the LPSZ Guaranteed Time has been spent to observe about three quaters of the sample. Qualitatively, preliminary analysis permit us to single out several interesting science case to study bimodality, merger events and shocks. A standard analysis method has been devised: at the map level, the calibration and map making use NIKA2 \emph{Baseline} calibration method~\cite{NIKA2-performance}, while the angular scale filtering induced by the data reduction is accounted for following~\cite{Ruppin2018}. A novel method, dubbed PANCO2, has been developped for the reconstruction of the thermodynamical profiles from the maps, and further developments are being conducted in view of the statistical analysis of the sample~\cite{Keruzore2021}.

\section{First results}
\label{Sect_results}

\begin{figure*}
\centering
\includegraphics[width=0.45\textwidth]{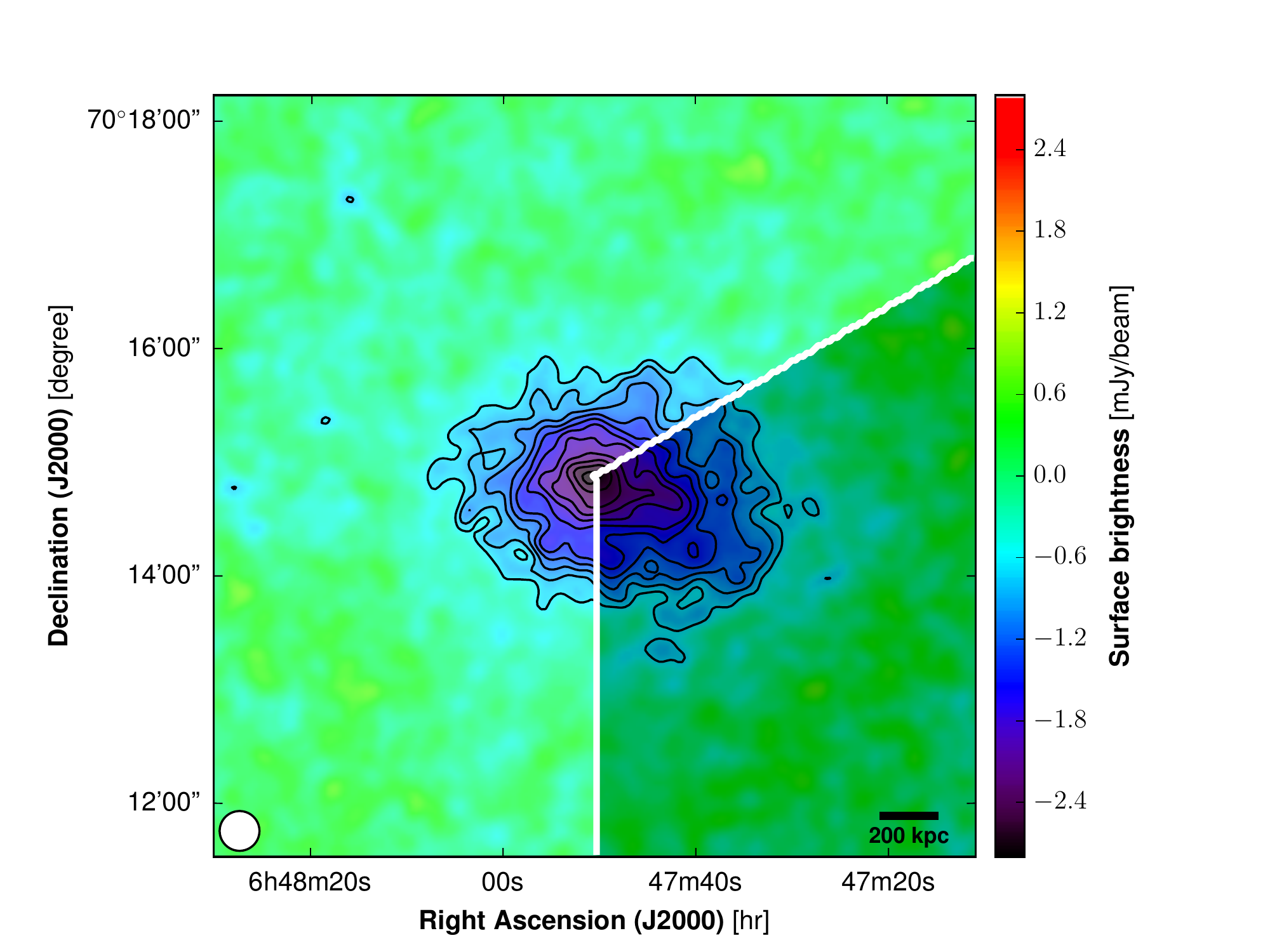}
\includegraphics[width=0.42\textwidth, clip=true, trim=-0.3cm 0cm 10cm 0cm]{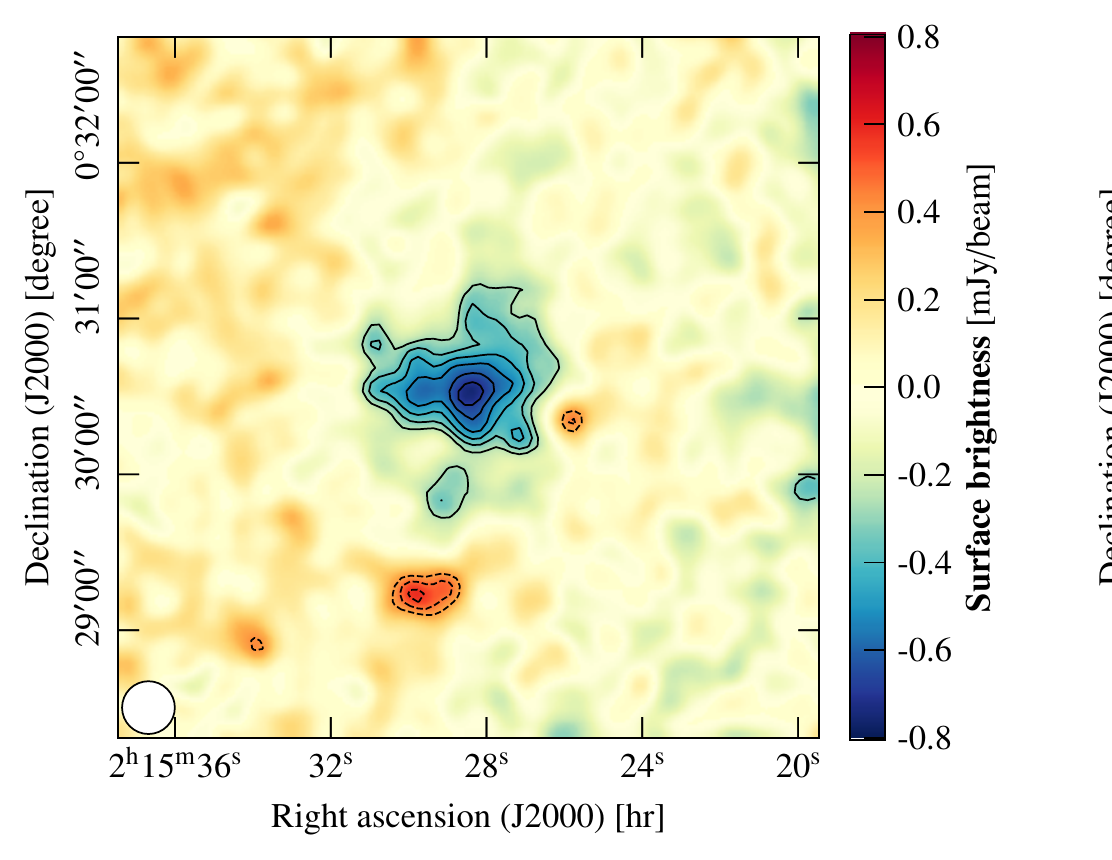}
\caption{First LPSZ results. Left: NIKA2 map at $150$-GHz towards the massive ($M_{500} = 7.8 \times 10^{14} M_{\odot}$) intermediate-redshift ($z \sim 0.6$) cluster PSZ2-G0144.83+25.11 after point source removal using NIKA2 $260$-GHz observation and ancillary data. The shaded sector shows the location of the over-pressure region that was evidenced as discussed in~\cite{Ruppin2018}; Right: NIKA2 $150$-GHz map of the low-mass ($M_{500} = 3.5 \times 10^{14} M_{\odot}$) high-redshift ($z \sim 0.9$) cluster ACT-CL J0215.4+0030 before point source treatment, as described in~\cite{Keruzore2020}. In both maps, contours show signal-to-noise levels starting from $3\sigma$ with a $1\sigma$ spacing.}
\label{Fig_maps}    
\end{figure*}

For the first LPSZ analysis, we chose PSZ2-G0144.83+25.11, a massive cluster of the first redshift-bin of the sample, making it an ideal target for observation as part of the Science Verification in April 2017. High-quality data were obtained enabling a $13.5~\sigma$ measurement of the SZ peak in the $150$-GHz map~\cite{Ruppin2018}. An over-pression region was evidenced, as shown with the shaded area in Fig.~\ref{Fig_maps} (left), which significatively impacts the pressure profile estimates inferred from the map. In turn, whether the over-pressure region is included or masked out affects the hydrostatic mass profile estimate, and finally the measured integrated Compton parameter and mass of the cluster. As discussed in~\cite{Ruppin2018}, this illustrates the capabilities of the LPSZ to assess the impact of the morphology on the mass-observable scaling relation.

By contrast, one of the most observationaly challenging cluster of the LPSZ sample was chosen as the object of the second LPSZ analysis: ACT-CL J0215.4+0030, a low-mass cluster in the highest redshift bin of the sample. While the SZ peak measurement reached $9\sigma$, the contour of $3\sigma$ signal-to-noise level in the map at $150$ GHz encloses only one square arcmin, as shown in Fig.~\ref{Fig_maps} (right). The compactness of this cluster required special treatment of the point sources using NIKA2 observation at 260~GHz and ancillary data, as described in~\cite{Keruzore2020}. Combining NIKA2, ACT and XMM-Newton data, the thermodynamical profiles of the cluster ICM were estimated, and the integrated Compton parameter and hydrostatic mass were inferred, improving the uncertainties on the hydrostatic mass measurements with respect to both ACT and XMM-Newton measurements~\cite{Keruzore2020}. This shows that the powerful synergetic SZ/X-ray method can be fully deployed even for the faintest and more compact cluster of the LPSZ.

\section{Conclusion}

Improving the reliability of the cosmological model inferred from galaxy clusters requires a better knowledge of cluster physics and in particular, an accurate estimate of their mass and their universal properties. 
%
%
However the universal pressure profile and the hydrostatic mass of clusters are estimated from X-ray observation of massive low-redshift cluster samples.
Using high-angular resolution observation of clusters in the millimeter-domain, we fully benefit from the redshift-independence property of the thermal SZ effect and from an observable directly proportional to the pressure of the ICM. Therefore combining X-ray and resolved thermal SZ observation of a representative sample of clusters is the key for extending to high redshift and low mass the estimation of the observable-to-mass scaling law and the universal pressure profile of clusters.   


NIKA2 is perfectly suited for this endavour as it combines high-angular resolution at 150 and 260 GHz with a large field-of-view, enabling esquisite resolved SZ mapping of clusters. This will be done in the framework of the LPSZ, a NIKA2 Guaranteed-Time Large Program dedicated to the observations of a representative sample of 45 SZ-selected clusters at $0.5 < z < 0.9$ from the Planck and ACT catalogs (300 hours, PI: F. Mayet, co-PI: L. Perotto).

The first LPSZ results demonstrate the synergetic capabilities of NIKA2 resolved SZ and XMM-\emph{Newton} X-ray observations for the in-depth study of the ICM thermal properties, including for the faintest more compact cluster of the sample, and hint at the future implications for Cosmology of the whole sample analysis. As main LPSZ products,  we will deliver the mean pressure profile and the scaling relation between the integrated Compton parameter and the mass, along with the surface brightness maps, the thermodynamical profiles and the codes that permits their inferrence. In addition to the X-ray/SZ analysis, we will perform multi-wavelength analysis combining LPSZ data with lensing, optical photometry/spectrometry, radio data to build a consistent picture of the cluster physics and further the LPSZ implication on cluster-based Cosmology.

\section*{Acknowledgements}
\small{We would like to thank the IRAM staff for their support during the campaigns. The NIKA2 dilution cryostat has been designed and built at the Institut N\'eel. In particular, we acknowledge the crucial contribution of the Cryogenics Group, and in particular Gregory Garde, Henri Rodenas, Jean Paul Leggeri, Philippe Camus. This work has been partially funded by the Foundation Nanoscience Grenoble and the LabEx FOCUS ANR-11-LABX-0013. This work is supported by the French National Research Agency under the contracts "MKIDS", "NIKA" and ANR-15-CE31-0017 and in the framework of the "Investissements d’avenir” program (ANR-15-IDEX-02). This work has benefited from the support of the European Research Council Advanced Grant ORISTARS under the European Union's Seventh Framework Programme (Grant Agreement no. 291294). F.R. acknowledges financial supports provided by NASA through SAO Award Number SV2-82023 issued by the Chandra X-Ray Observatory Center, which is operated by the Smithsonian Astrophysical Observatory for and on behalf of NASA under contract NAS8-03060.}

%
%
%

\end{document}